\documentclass[aps,prl,twocolumn,superscriptaddress,groupedaddress]{revtex4-1}  
\pdfoutput=1
\usepackage{graphicx}  
\usepackage{dcolumn}   
\usepackage{bm}        
\usepackage{amssymb}   
\usepackage{amsmath}
\usepackage{gensymb}
\usepackage{subfigure}
\begin{document}
	\title{Measurement of the radial mode spectrum of photons through a phase-retrieval method}
	
	\author{Saumya Choudhary}
	\email{schoudha@ur.rochester.edu}
	\affiliation{The Institute of Optics, University of Rochester, Rochester, NY 14627, USA}
	\author{Rachel Sampson}
	\affiliation{CREOL, The College of Optics and Photonics at UCF, Orlando, FL 32816, USA}
	\author{Yoko Miyamoto}
	\affiliation{Department of Engineering Science, The University of Electro-Communications, 1-5-1 Chofugaoka, Chofu, Tokyo 182-8585, Japan}
	\author{Omar S. Maga\~na-Loaiza}
	\affiliation{Department of Physics and Astronomy, Louisiana State University, Baton Rouge, LA 70803, USA}
	\author{Seyed Mohammad Hashemi Rafsanjani}
    \affiliation{Department of Physics, University of Miami, Coral Gables, Florida 33146, USA}
	\author{Mohammad Mirhosseini}
    \affiliation{Department of Applied Physics and Material Science, California Institute of Technology, Pasadena, CA 91125, USA}
	\author{Robert W. Boyd}
	\affiliation{The Institute of Optics, University of Rochester, Rochester, NY 14627, USA}
	\affiliation{Department of Physics, University of Ottawa, Ottawa, ON K1N 6N5, Canada}

\begin{abstract}
	We propose and demonstrate a simple and easy-to-implement projective-measurement protocol to determine the radial index $p$ of a Laguerre-Gaussian ($LG^{l}_p$) mode. Our method entails converting any specified high-order $LG^{0}_p$ mode into a near-Gaussian distribution that matches the fundamental mode of a single-mode fiber (SMF) through the use of two phase-screens (unitary transformations) obtained by applying a phase-retrieval algorithm. The unitary transformations preserve the orthogonality of modes and guarantee that our protocol can, in principle, be free of crosstalk. We measure the coupling efficiency of the transformed radial modes to the SMF for different pairs of phase-screens. Because of the universality of phase-retrieval methods, we believe that our protocol provides an efficient way of fully characterizing the radial spatial profile of an optical field.
\end{abstract}
\date{\today}
\maketitle

\indent Laguerre-Gaussian (LG) modes, $LG^{l}_p$, characterized by the radial index $p$ (a non-negative integer) and by a azimuthal index $l$ (an integer), are solutions of the paraxial wave equation in cylindrical coordinates \cite{Milonni, Allen}. The LG modes are orthonormal and form a complete basis set. Both the azimuthal and radial degrees of freedom are theoretically unbounded (unlike polarization, which is 2-dimensional). This unbounded nature is potentially useful for applications such as classical and quantum communication \cite{Communication, Willner, Willner2, Willner3, Gibson}, including quantum key distribution (QKD) \cite{QKD, Sit} by allowing for high-dimensional encoding, which leads to increased information capacity \cite{Cerf, Tomography1}. To date, most of these applications have employed the $LG^{l}_0$, that is, the OAM modes of lowest radial order, for which efficient means of detection and characterization are available \cite{Boyd_sorter, Padgett_sorter, Leach, qplate, Integrated_sorter, Zeilinger_projection}. The radial modes, $LG^{l}_p$, have been under-utilized for such applications because of the absence of methods for detecting and measuring the radial index.\\
\indent  Some methods for measuring the radial index have recently been reported \cite{Yiyu, Ebrahim, Fickler, Willner4}. Based on the principle of a universal quantum sorter \cite{Radu}, Zhou et al. \cite{Yiyu} demonstrated  the sorting of even- and odd-order radial modes into separate output ports of an interferometer. However, further classification of a large number of modes would require cascading several such interferometers, thereby increasing the implementation complexity. Projective measurement presents one method for mode-sorting, and generally entails mapping a specified input mode to a particular output mode (typically a Gaussian). This output mode can then be coupled to a SMF that selects only the fundamental Gaussian component. The projective measurement method for radial modes demonstrated in \cite{Ebrahim} employs flattening the phase-front of the incoming field. However, the incomplete projection onto the SMF mode results in low (input mode-dependent) detection efficiency as well as crosstalk, which are detrimental for most quantum applications. Although the crosstalk can be reduced by selecting the central (flat intensity) portion of the phase-flattened mode, it would also decrease the detection efficiency \cite{Bouchard}. \\
\indent LG modes retain their transverse structure as they propagate through free space, through lenses and as they reflect off of mirrors. As a consequence, one cannot use these elements to losslessly project high-order radial modes onto a Gaussian. Morizur et al. proposed \cite{Treps} that unitary transformation of any given mode to another can be achieved by repeated reflection off a deformable mirror (which introduces a spatially varying transverse phase) and Fourier transformation through a lens, over several iterations. However, their protocol was only tested for the first four Hermite-Gaussian modes and required several optimization steps to arrive at the correct configuration of the deformable mirror required for high conversion efficiency. Repeated phase transformations introduced by either reflecting off a spatial light modulator (SLM) or by multiple scattering have been used in \cite{Fontaine} and in \cite{Fickler} respectively, to map an LG mode to a mode index-dependent position at the output. However, while the former configuration cannot be used to extract the relative phases between constituent modes in a coherent superposition, the latter suffers from high scattering losses (approximately $0.2\%$ of input power was collected at the output).\\
\indent  Here, we present a simple and easy-to-implement protocol for measuring the radial mode spectrum of an optical field by transforming high-order radial modes to the fundamental mode of a SMF. In our approach, we use two phase-screens (to introduce a spatially varying transverse phase structure), placed at the object plane and at the Fourier plane of a lens respectively, to transform a particular radial mode ($LG^0_p$) to the fundamental mode of a SMF, assumed here to be a Gaussian ($LG^0_0$). The required pair of phase-screens for each input mode is calculated through the use of a phase-retrieval algorithm. The transformations performed by phase-screens are in principle lossless and hence, unitary. The orthogonality of modes is maintained after unitary transformation, and this property guarantees that modes with radial index other than that for which the phase screens were designed will not couple to the SMF. Thus, through our protocol, one can measure the radial index of a mode (and thereby determine the radial mode spectrum of an optical field) with negligible crosstalk.
\begin{figure}[ht!]
	\centering
	\includegraphics[width=78mm]{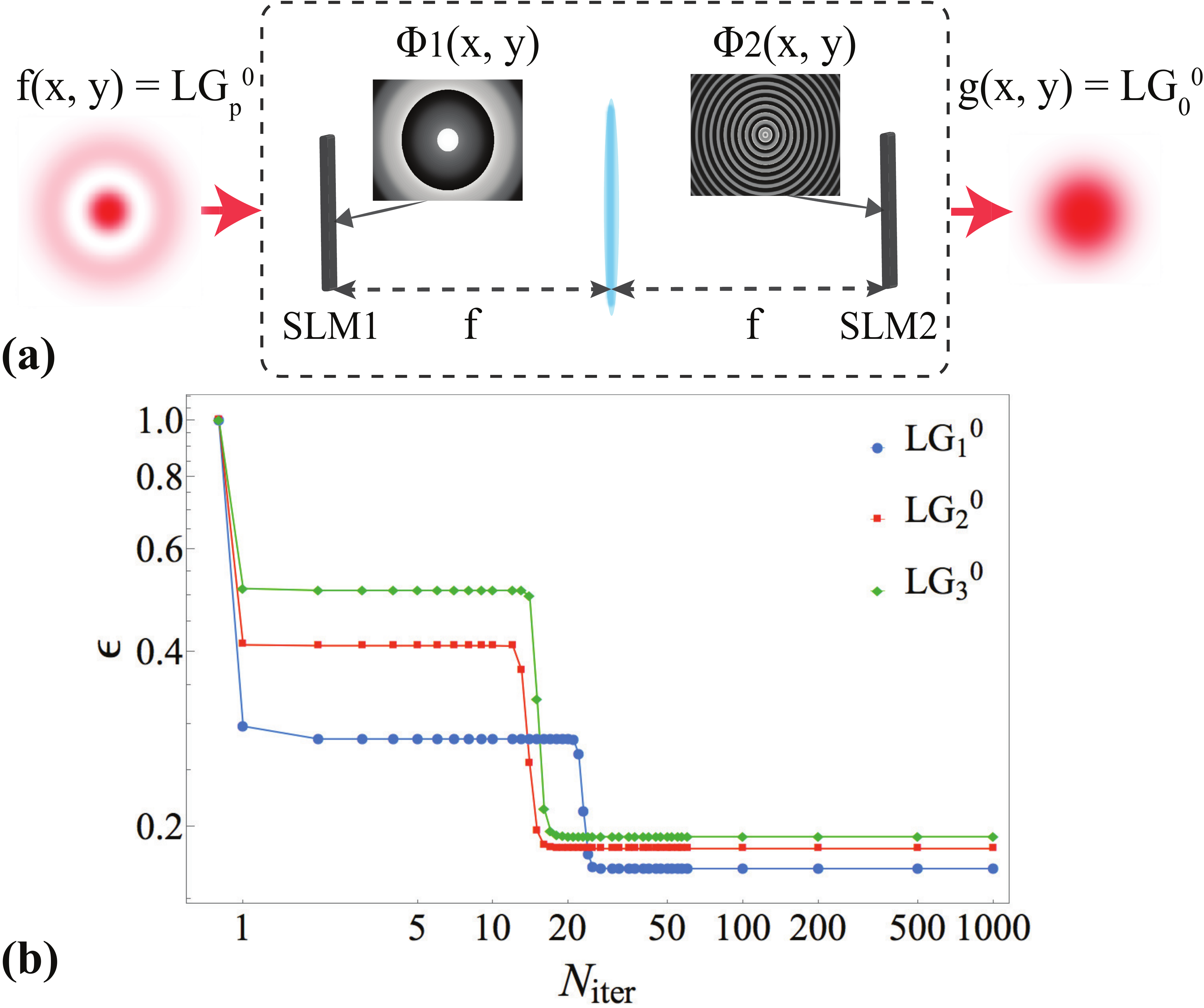}
	\caption{(a) Schematic representation of the protocol with insets showing the phase-screens, $\Phi_1(x, y)$ and $\Phi_2(x, y)$ that convert an $LG^0_p$ mode into a Gaussian output mode. (b) The variation of Fourier-domain error $\epsilon$ for the input modes $LG^0_1$, $LG^0_2$ and $LG^0_3$, with the number of iterations $N_{\text{iter}}$ of the algorithm.}
	\label{fig:conceptual}
\end{figure}

\begin{figure}[ht!]
	\centering
	\includegraphics[width=85mm]{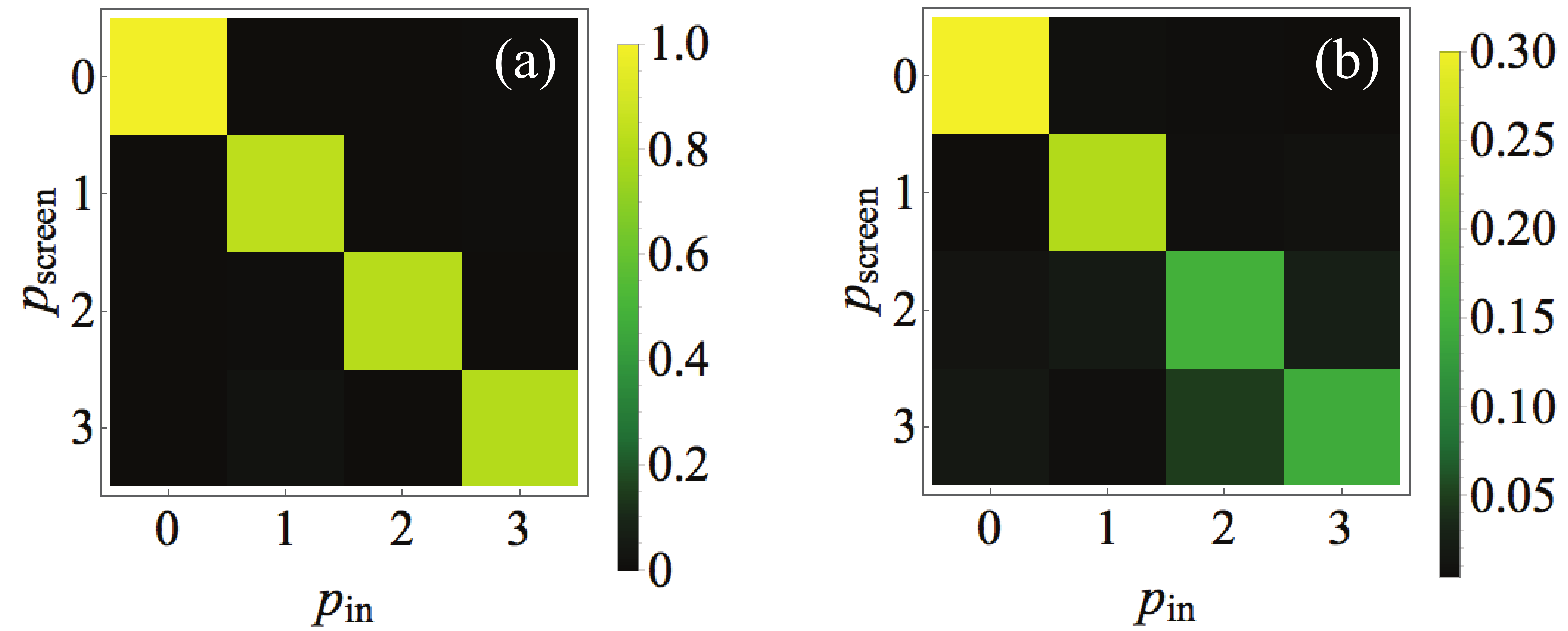}
	\caption{Crosstalk matrices (a) from calculations, and (b) obtained experimentally. Here $p_{\text{in}}$ is the radial index of the input mode, whereas $p_{\text{screen}}$ is the radial index corresponding to the pair of phase-screens. See tables~\ref{table:calculated} and~\ref{table:experiment} in the appendix for the actual values.}
	\label{fig:crosstalk}
\end{figure}

\indent Figure~\ref{fig:conceptual}(a) shows a schematic representation of our projection protocol. The transformation of a $LG^0_{p}$ mode (where $p>0$) to a Gaussian mode is an example of a `synthesis problem' as we have \textit{a priori} knowledge of the intensity distributions in the object plane as well as the Fourier plane \cite{Fienup2}. We use the Gerchberg-Saxton (GS) phase-retrieval algorithm, an iterative error-reduction algorithm, \cite{Gerchberg, Fienup} to calculate the required phase-screens $\Phi_1(x, y)$ and $\Phi_2(x, y)$. The SLM1, placed at the object plane of a lens, introduces a phase $\Phi_1(x, y)$ on the input field $LG^0_p$ (or $f(x, y)$). On propagation through the lens, the intensity distribution at the Fourier plane (or equivalently at the far-field) becomes similar to the intensity distribution of the $LG^0_0$ mode (or $g(x, y)$). However, this transformation often results in an incorrect phase distribution at the Fourier plane. The SLM2, placed at the Fourier plane, subsequently introduces another phase, $\Phi_2(x, y)$, which corrects these residual phase errors and flattens the phase. As phase-transformations in the absence of losses are unitary, the orthogonality of these projected modes remains preserved at the SMF. Due to restrictions on the aperture size of our SLM, we choose the first three high-order radial modes $LG^0_p$ with $p = \{1, 2, 3\}$ to test our protocol.\\
\indent For our two-dimensional problem, the uniqueness of a phase-retrieval solution is ensured due to the non-factorability of polynomials of two-or-more complex variables \cite{Fienup_unique}. We define an error metric $\epsilon$ in the Fourier domain to study the convergence of the GS algorithm to this solution with the number of iterations $N_{\text{iter}}$. In many cases, if a solution exists, the algorithm converges within a few iterations. However, the algorithm can also stagnate close to the local minimum of $\epsilon$ without converging any further \cite{Fienup2}. To study the efficiency of the algorithm at calculating the required pair of phase-screens for each input mode, we define $\epsilon$ in terms of $C$, the coherent mode-overlap efficiency between the projected $LG^0_p$ mode and the mode of the SMF, as follows
\begin{equation}\label{eq:error}
\epsilon = 1 - C = 1 - |2\pi \sum_{\rho} \rho F^*_n(\rho) g_n(\rho)|^2.
\end{equation}
Here, $\rho$ is the Fourier domain radial coordinate, $F_n(\rho)$ is the normalized radial mode at the Fourier plane after the application of the relevant pair of phase-screens $\Phi_1$ and $\Phi_2$, and $g_n(\rho)$ is the desired mode at the Fourier plane (normalized Gaussian). The modes are normalized such that $2\pi \sum_{\rho} (\rho |F_n(\rho)|^2) = 2\pi \sum_{\rho} (\rho |g_n(\rho)|^2) = 1$. $C$ also represents the fraction of input power that couples into the SMF after the phase-corrections. Figure~\ref{fig:conceptual}(b) shows the variation of $\epsilon$ with $N_{\text{iter}}$, where $N_{\text{iter}}$ is the number of iterations, for the different input radial modes ($LG^0_1$ to $LG^0_3$). Without the phase-corrections, or when $N_{\text{iter}} = 0$, $\epsilon$ is unity as the unconverted high-order radial modes are orthogonal to the SMF mode. As $N_{\text{iter}}$ increases, $\epsilon$ ($C$) decreases (increases) for all modes until it stagnates close to 0.18 after approximately 35 iterations. This non-negligible error represents an overall loss of power coupled into the SMF. However, as we show later in the paper, this loss of power is not as consequential as there is negligible crosstalk due to the preserved orthogonality of modes in our protocol (see figure~\ref{fig:crosstalk}(a)).
\begin{figure}[ht!]
	\centering
	\includegraphics[width=72mm]{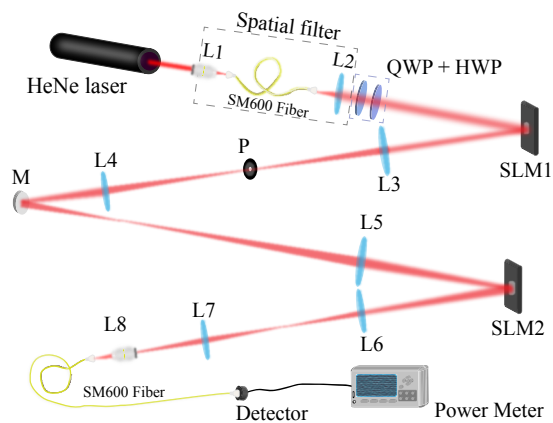}
	\caption{Schematic of the experimental setup. The labeled components are: Collimating lens (L2), microscope objectives (L1 and L8) to couple laser light out of and into a SMF respectively, waveplates (QWP+HWP) to transform the polarization from the output of the fiber to the correct polarization for SLM1, pupil P (at the Fourier plane of lens L3) to isolate the first order of diffraction from SLM1, imaging systems to create a magnified image of the Fourier plane of L3 onto SLM2 (lenses L4 and L5) and to image the transformed field onto the exit pupil of L8 (lenses L6 and L7).}
	\label{fig:Schematic}
\end{figure}

\begin{figure}[ht!]
	\centering
	\includegraphics[width=78mm]{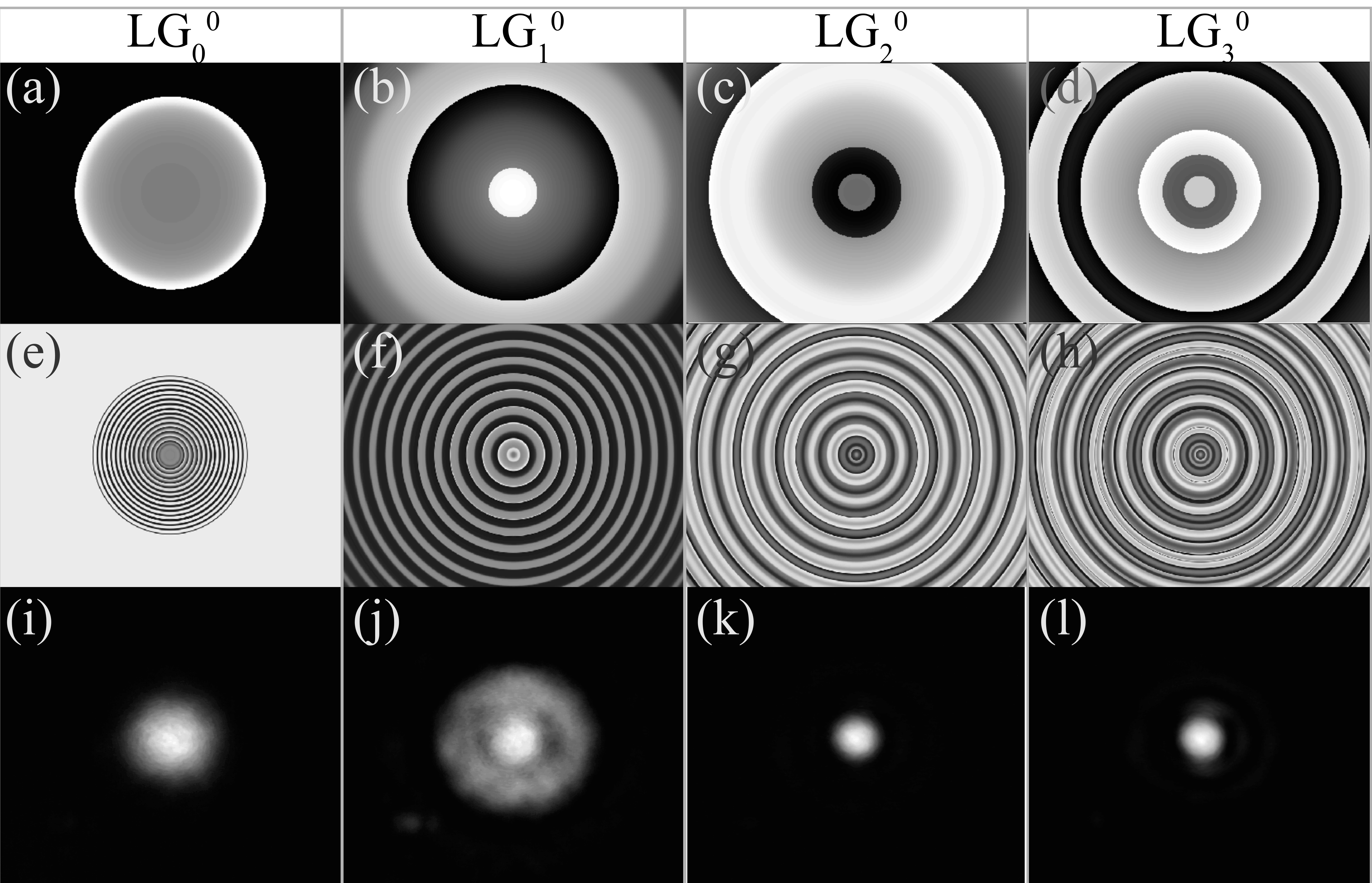}
	\caption{(a-d) Phase-screens on SLM1, and (e-h) on SLM2 for input radial modes $LG^0_0$ to $LG^0_3$ (from left to right). In calculating these phase screens, GS algorithm was applied for 200 iterations (well beyond the stagnation point). (j-l) Captured intensities at SLM2 after applying the corresponding phase-screens to the various input modes $LG^0_0$ to $LG^0_3$ (from left to right). In all cases, the transformed intensity closely resembles the near-Gaussian mode of a SMF, although some distortion is evident in panel (j).}
	\label{fig:Screens}
\end{figure}

\indent Figure~\ref{fig:Schematic} shows a schematic of the experimental setup used to test the protocol. Two Hamamatsu LCOS (Liquid crystal on Silicon) SLMs (model X10468-02), with a pixel size of 20 $\mu m$ and an active area of 600 $\times$ 800 pixels, are used as the phase screens. The source is a He-Ne laser spatially filtered through the use of a SM600 fiber (single mode for 633 nm wavelength). The Gaussian mode at the output of the SMF is collimated to a large diameter to ensure that the beam is larger than the active area of our SLMs to have uniform intensity at all the pixels of SLM1. The $LG^0_p$ modes ($p = \{0, 1, 2, 3\}$) are generated by phase-only modulation of SLM1 through the use of computer-generated holograms. The amplitude and phase of the LG mode is imposed on the first order of diffraction of the grating function impressed on SLM1, in accordance with the approach given in \cite{Gonzalez}. For convenience and efficient resource utilization, the first phase-screen (see figures~\ref{fig:Screens}(a)-(d)) is implemented along with the mode generation on SLM1 itself. The azimuthally-averaged intensity distributions for each input radial mode at the image plane and the Fourier plane, recorded using a CCD camera, were found to differ from the theoretical intensities by a root-mean-square (rms) error of less than $10 \%$ (see figure~\ref{fig:Fidelity} in the appendix), thereby confirming that the generated radial modes are of good fidelity. After the second phase-correction impressed by SLM2 (see figures~\ref{fig:Screens}(e)-(h)), the transformed modes are coupled via an imaging system and a microscope objective to a SMF-coupled photodetector. Figures~\ref{fig:Screens}(i)-(l) show the recorded intensities at the Fourier plane (or SLM2) of the various input modes after the application of the corresponding pair of phase-screens. 

The coupling efficiency of a given mode to the SMF is taken to be the ratio of the power coupled into the fiber to the total power incident on the microscope objective. This method of calculation allows us to account for any losses prior to the objective. The coupling efficiencies of different input radial modes (from $LG^0_0$ to $LG^0_3$) for various pairs of phase-screens are obtained similarly, forming the `crosstalk matrix'. We note that although the $LG^0_0$ mode simply requires a rescaling of the waist size for efficient coupling to the SMF, we include it in our basis set for completeness and for examining the crosstalk.
\begin{figure}[ht!]
	\centering
	\includegraphics[width=78mm]{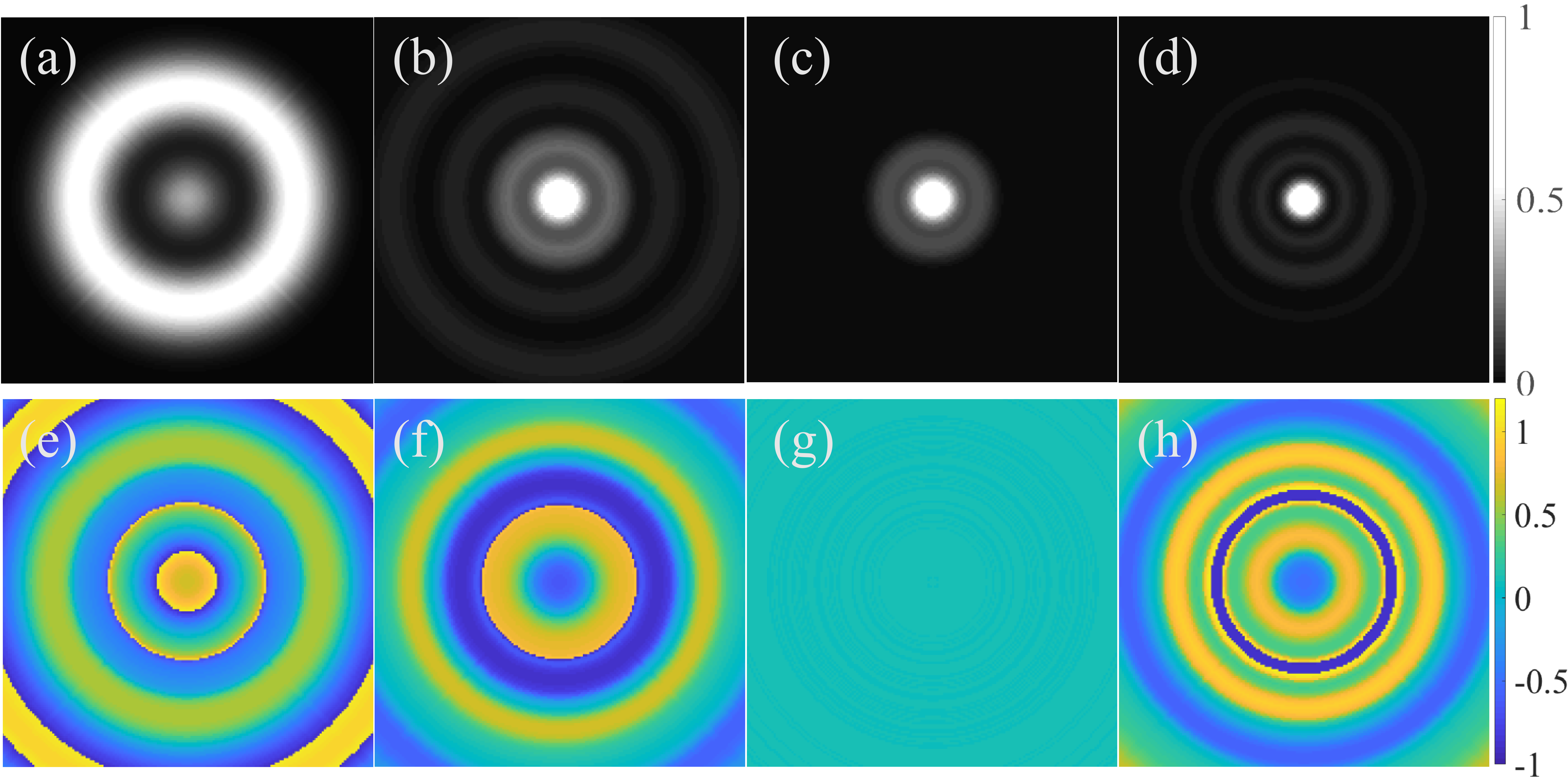}
	\caption{Calculated (a-d) intensities and (e-h) phases of input modes $LG^0_0$ to $LG^0_3$ (from left to right) at the Fourier plane after the application of phase-corrections corresponding to $LG^0_2$ mode. The intensities are normalized to the maximum intensity and the phases are given in units of $\pi$ radians. Note that only when the input mode corresponds to the phase-screens on the two SLMs does the transformed mode have the correct amplitude (c) and (flat) phase (g) distributions to couple to the Gaussian mode profile of a single-mode fiber. }
	\label{fig:pscreen2}
\end{figure}

Figure~\ref{fig:pscreen2} shows the calculated intensities and phase-fronts of different input modes ($LG^0_0$, $LG^0_1$, $LG^0_2$ and $LG^0_3$) at the Fourier plane after the application of phase-screens calculated for the $LG^0_2$ mode. Essentially, one obtains a nearly-Gaussian amplitude and a flat-phase at the Fourier plane only when the applied phase-screens correspond to the particular input mode. Since the coupling efficiency of the transformed modes to the SMF depends on both intensity and phase distributions, the improperly transformed modes have negligible coupling efficiency to the SMF (or equivalently, negligible crosstalk). Hence, the coherent mode-overlap efficiency, or $C$ in equation~\ref{eq:error}, between the transformed modes and the SMF mode is the pertinent measure for the efficacy of these phase-screens.

\indent Figure~\ref{fig:crosstalk}(a) shows the predicted crosstalk matrix, which is obtained by calculating the overlap integral of each transformed radial mode with the SMF mode for different pairs of phase-screens. The diagonal elements of the matrix have high values (larger than 0.8), implying a high conversion efficiency for the radial modes with the correct pair of phase-screens. The off-diagonal elements, on the other hand, have low values (close to zero), implying a negligible conversion efficiency for the incorrect pair of phase-screens. Further, in agreement with figure~\ref{fig:conceptual}(b), the conversion efficiency of radial modes (diagonal elements of figure~\ref{fig:crosstalk}(a)) decreases monotonically with increasing radial index, from 99$\%$ for $LG^0_0$ to approximately 81$\%$ for $LG^0_3$. Figure~\ref{fig:crosstalk}(b) shows the measured crosstalk matrix. Comparing the diagonal terms of the matrices in figures~\ref{fig:crosstalk}(b) and ~\ref{fig:crosstalk}(a), we see that the coupling efficiencies for all the modes, with the correct pair of phase-screens is lower than the calculated result. Also, except for $LG^0_3$ (or $p_{\rm{in}} = 3$), the measured crosstalk values, given by the off-diagonal terms, are less than $15 \%$. \\
\indent The lower coupling efficiency measured overall, and the observed crosstalk for radial indices larger than 1 could be due to a combined effect of the following factors: (1) calibration errors at each pixel on the SLMs (Correct calibration of each pixel is a stringent requirement for SLM2. SLM1 has a grating for mode generation. Therefore, any calibration errors therein would manifest in the diffraction efficiency, and not as significantly in the profile of the generated mode.); (2) imperfect optics, and the presence of aberrations such as astigmatism and spherical aberration in the imaging systems before and after SLM2; (3) crosstalk between the pixels on the SLMs (called the fringing effect in \cite{fringing}), which becomes significant whenever the phase wraps from $2\pi$ to 0 gradually over a few pixels instead of sharply. A possible solution for improving the coupling efficiency and lowering the crosstalk could be to use a genetic algorithm similar to the one used in \cite{Fickler} to calibrate and correct for the phase errors due to the SLMs as well as the imaging systems. Also, a hologram with high spatial resolution should reduce the crosstalk between adjacent pixels during phase-wrapping. \\
\indent To summarize, we have proposed and provided a proof-of-principle demonstration of a new protocol for determining the radial mode decomposition of an optical field. The protocol utilizes two phase transformations, one at the object and the other in the Fourier plane of a lens, to convert high-order radial modes to the fundamental mode of a SMF. The required phase transformations were calculated using the Gerchberg-Saxton phase-retrieval algorithm. The implementation is straightforward and does not require complicated setups and optimization. Also, as it utilizes only phase-corrections, our procedure for maximizing the coherent mode overlap between the high-order LG modes and the Gaussian mode is intrinsically non-lossy. We believe that by improving on the implementation, as suggested in the preceding paragraph, the performance can be improved further.\\
\indent The universal nature of the phase-retrieval algorithm suggests the future use of this protocol for measurements of the azimuthal mode structure in addition to the radial modes, or for measurements in other bases including the Hermite-Gaussian basis. This possibility opens up a plethora of applications pertaining to classical and quantum communication, and to quantum computation \cite{Chuang} that benefit from the increased information capacity obtained by accessing the entire transverse degree of freedom of photons. In addition, this protocol may be useful for applications such as super-resolution \cite{Tsang} and quantum state tomography \cite{Tomography1, Tomography2}.\\
\indent The authors thank N. Treps, Y. Zhou, J. Zhao, T. Gerrits and L. Hernandez for helpful discussions. The authors also acknowledge funding support from the U.S. Office of Naval Research (grant number: ONR N00014-17-1-2443). Y. M. acknowledges funding support from JSPS KAKENHI (grant number: JP16K05499)

\bibliographystyle{unsrt}
\bibliography{Ppr_refs_new}

\begin{thebibliography}{10}

\bibitem{Milonni}
Peter~W. Milonni and Joseph~H. Eberly.
\newblock {\em Laser Physics}, pages 269--329.
\newblock John Wiley \& Sons, Inc., 2010.

\bibitem{Allen}
L.~Allen, M.~W. Beijersbergen, R.~J.~C. Spreeuw, and J.~P. Woerdman.
\newblock Orbital angular momentum of light and the transformation of
  laguerre-gaussian laser modes.
\newblock {\em Phys. Rev. A}, 45:8185--8189, Jun 1992.

\bibitem{Communication}
J.~Wang, J.~Y. Yang, I.~M Fazal, N.~Ahmed, Y.~Yan, H.~Huang, Y.~Ren, Y.~Yue,
  S.~Dolinar, M.~Tur, et~al.
\newblock Terabit free-space data transmission employing orbital angular
  momentum multiplexing.
\newblock {\em Nat. Phot.}, 6(7):488--496, 2012.

\bibitem{Willner}
N.~Bozinovic, Y.~Yue, M.~Ren, Y.and~Tur, P.~Kristensen, H.~Huang, A.~E.
  Willner, and S.~Ramachandran.
\newblock Terabit-scale orbital angular momentum mode division multiplexing in
  fibers.
\newblock {\em Science}, 340(6140):1545--1548, 2013.

\bibitem{Willner2}
A.~E. Willner, H.~Huang, Y.~Yan, Y.~Ren, N.~Ahmed, G.~Xie, C.~Bao, L.~Li,
  Y.~Cao, Z.~Zhao, J.~Wang, M.~P.~J. Lavery, M.~Tur, S.~Ramachandran, A.~F.
  Molisch, N.~Ashrafi, and S.~Ashrafi.
\newblock Optical communications using orbital angular momentum beams.
\newblock {\em Adv. Opt. Photon.}, 7(1):66--106, Mar 2015.

\bibitem{Willner3}
G.~Xie, Y.~Ren, Y.~Yan, H.~Huang, N.~Ahmed, L.~Li, Z.~Zhao, C.~Bao, M.~Tur,
  S.~Ashrafi, and A.~E. Willner.
\newblock Experimental demonstration of a 200-gbit/s free-space optical link by
  multiplexing laguerre--gaussian beams with different radial indices.
\newblock {\em Optics letters}, 41(15):3447--3450, 2016.

\bibitem{Gibson}
G.~Gibson, J.~Courtial, M.~J. Padgett, M.~Vasnetsov, V.~Pas'ko, S.~M. Barnett,
  and S.~Franke-Arnold.
\newblock Free-space information transfer using light beams carrying orbital
  angular momentum.
\newblock {\em Opt. Exp.}, 12(22):5448--5456, Nov 2004.

\bibitem{QKD}
M.~Mirhosseini, O.~S. Maga{\~n}a-Loaiza, M.~N O'Sullivan, B.~Rodenburg,
  M.~Malik, M.~P.~J Lavery, M.~J. Padgett, D.~J Gauthier, and R.~W. Boyd.
\newblock High-dimensional quantum cryptography with twisted light.
\newblock {\em N. J. of Phys.}, 17(3):033033, 2015.

\bibitem{Sit}
Alicia Sit, Fr\'{e}d\'{e}ric Bouchard, Robert Fickler, J\'{e}r\'{e}mie
  Gagnon-Bischoff, Hugo Larocque, Khabat Heshami, Dominique Elser, Christian
  Peuntinger, Kevin G\"{u}nthner, Bettina Heim, Christoph Marquardt, Gerd
  Leuchs, Robert~W. Boyd, and Ebrahim Karimi.
\newblock High-dimensional intracity quantum cryptography with structured
  photons.
\newblock {\em Optica}, 4(9):1006--1010, Sep 2017.

\bibitem{Cerf}
Mohamed Bourennane, Anders Karlsson, Gunnar Björk, Nicolas Gisin, and
  Nicolas~J Cerf.
\newblock Quantum key distribution using multilevel encoding: security
  analysis.
\newblock {\em Journal of Physics A: Mathematical and General}, 35(47):10065,
  2002.

\bibitem{Tomography1}
Mario Krenn, Marcus Huber, Robert Fickler, Radek Lapkiewicz, Sven Ramelow, and
  Anton Zeilinger.
\newblock Generation and confirmation of a (100 x 100)-dimensional entangled
  quantum system.
\newblock {\em Proceedings of the National Academy of Sciences},
  111(17):6243--6247, 2014.

\bibitem{Boyd_sorter}
M.~Mirhosseini, M.~Malik, Z.~Shi, and R.~W. Boyd.
\newblock Efficient separation of the orbital angular momentum eigenstates of
  light.
\newblock {\em Nat. Comm.}, 4, 2013.

\bibitem{Padgett_sorter}
G.~C.~G Berkhout, M.~P.~J. Lavery, J.~Courtial, M.~W. Beijersbergen, and M.~J.
  Padgett.
\newblock Efficient sorting of orbital angular momentum states of light.
\newblock {\em Phys. Rev. Lett.}, 105(15):153601, 2010.

\bibitem{Leach}
J.~Leach, J.~Courtial, K.~Skeldon, S.~M. Barnett, S.~Franke-Arnold, and M.~J.
  Padgett.
\newblock Interferometric methods to measure orbital and spin, or the total
  angular momentum of a single photon.
\newblock {\em Phys. Rev. Lett.}, 92(1):013601, 2004.

\bibitem{qplate}
E.~Karimi, B.~Piccirillo, E.~Nagali, L.~Marrucci, and E.~Santamato.
\newblock Efficient generation and sorting of orbital angular momentum
  eigenmodes of light by thermally tuned q-plates.
\newblock {\em App. Phys. Lett.}, 94(23):231124, 2009.

\bibitem{Integrated_sorter}
T.~Su, R.~P. Scott, S.~S. Djordjevic, N.~K. Fontaine, D.~J. Geisler, X.~Cai,
  and S.~J.~B. Yoo.
\newblock Demonstration of free space coherent optical communication using
  integrated silicon photonic orbital angular momentum devices.
\newblock {\em Opt. Exp.}, 20(9):9396--9402, 2012.

\bibitem{Zeilinger_projection}
A.~Mair, A.~Vaziri, G.~Weihs, and A.~Zeilinger.
\newblock Entanglement of the orbital angular momentum states of photons.
\newblock {\em Nature}, 412(6844):313--316, 2001.

\bibitem{Yiyu}
Yiyu Zhou, Mohammad Mirhosseini, Dongzhi Fu, Jiapeng Zhao, Seyed~Mohammad
  Hashemi~Rafsanjani, Alan~E. Willner, and Robert~W. Boyd.
\newblock Sorting photons by radial quantum number.
\newblock {\em Phys. Rev. Lett.}, 119:263602, Dec 2017.

\bibitem{Ebrahim}
H.~Qassim, F.~M. Miatto, J.~P. Torres, M.~J. Padgett, E.~Karimi, and R.~W.
  Boyd.
\newblock Limitations to the determination of a laguerre-gauss spectrum via
  projective, phase-flattening measurement.
\newblock {\em J. Opt. Soc. Am. B}, 31(6):A20--A23, Jun 2014.

\bibitem{Fickler}
R.~Fickler, M.~Ginoya, and R.~W. Boyd.
\newblock Custom-tailored spatial mode sorting by controlled random scattering.
\newblock {\em Phys. Rev. B}, 95:161108, Apr 2017.

\bibitem{Willner4}
G.~Xie, Y.~Ren, H.~Huang, N.~Ahmed, L.~Li, Y.~Yan, M.~Lavery, M.~Padgett, and
  A.~E. Willner.
\newblock In {\em Frontiers in Optics}, pages FTh4B--6. Optical Society of
  America, 2014.

\bibitem{Radu}
Radu Ionicioiu.
\newblock Sorting quantum systems efficiently.
\newblock {\em Scientific Reports}, 6:25356 EP --, 05 2016.

\bibitem{Bouchard}
F.~Bouchard, N.~H. Valencia, F.~Brandt, R.~Fickler, M.~Huber, and M.~Malik.
\newblock Measuring azimuthal and radial modes of photons.
\newblock {\em arXiv preprint arXiv:1808.03533}, 2018.

\bibitem{Treps}
J.~F. Morizur, L.~Nicholls, P.~Jian, S.~Armstrong, N.~Treps, B.~Hage, M.~Hsu,
  W.~Bowen, J.~Janousek, and H.A. Bachor.
\newblock Programmable unitary spatial mode manipulation.
\newblock {\em J. of Opt. Soc. of Am. A}, 27(11):2524--2531, 2010.

\bibitem{Fontaine}
N.~K. Fontaine, R.~Ryf, H.~Chen, D.T. Neilson, K.~Kim, and J.~Carpenter.
\newblock Optical spatial mode sorter of azimuthal and radial components.
\newblock {\em arXiv preprint arXiv:1803.04126}, 2018.

\bibitem{Fienup2}
J.~R. Fienup.
\newblock Reconstruction and synthesis applications of an iterative algorithm.
\newblock In {\em Transformations in Optical Signal Processing}, volume 373,
  pages 147--161. International Society for Optics and Photonics, 1984.

\bibitem{Gerchberg}
R.~W. Gerchberg and W.~O. Saxton.
\newblock {A practical algorithm for the determination of the phase from image
  and diffraction plane pictures}.
\newblock {\em Optik (Jena)}, 35:237, 1972.

\bibitem{Fienup}
J.~R. Fienup.
\newblock Phase retrieval algorithms: a comparison.
\newblock {\em App. Opt.}, 21(15):2758--2769, 1982.

\bibitem{Fienup_unique}
J.H. Seldin and J.~R. Fienup.
\newblock Numerical investigation of the uniqueness of phase retrieval.
\newblock {\em JOSA A}, 7(3):412--427, 1990.

\bibitem{Gonzalez}
V.~Arriz{\'o}n, U.~Ruiz, R.~Carrada, and L.~A. Gonz{\'a}lez.
\newblock Pixelated phase computer holograms for the accurate encoding of
  scalar complex fields.
\newblock {\em J. of Opt. Soc. of Am. A}, 24(11):3500--3507, 2007.

\bibitem{fringing}
T.~Lu, M.~Pivnenko, B.~Robertson, and D.~Chu.
\newblock Pixel-level fringing-effect model to describe the phase profile and
  diffraction efficiency of a liquid crystal on silicon device.
\newblock {\em Applied optics}, 54(19):5903--5910, 2015.

\bibitem{Chuang}
M.~A. Nielsen and I.~L. Chuang.
\newblock {\em Quantum computation and quantum information}.
\newblock Cambridge university press, 2010.

\bibitem{Tsang}
M.~Tsang, R.~Nair, and X.~M. Lu.
\newblock Quantum theory of superresolution for two incoherent optical point
  sources.
\newblock {\em Physical Review X}, 6(3):031033, 2016.

\bibitem{Tomography2}
B~Jack, J~Leach, H~Ritsch, SM~Barnett, MJ~Padgett, and S~Franke-Arnold.
\newblock Precise quantum tomography of photon pairs with entangled orbital
  angular momentum.
\newblock {\em N. J. of Phys.}, 11(10):103024, 2009.

\bibitem{Krenn}
William~N. Plick and Mario Krenn.
\newblock Physical meaning of the radial index of laguerre-gauss beams.
\newblock {\em Phys. Rev. A}, 92:063841, Dec 2015.

\end{thebibliography}

\onecolumngrid
\appendix
\newpage
\section*{Appendix}
\subsection{Fidelity of generated radial modes}  
\begin{figure}[ht!]
	\centering
	\includegraphics[width=110mm]{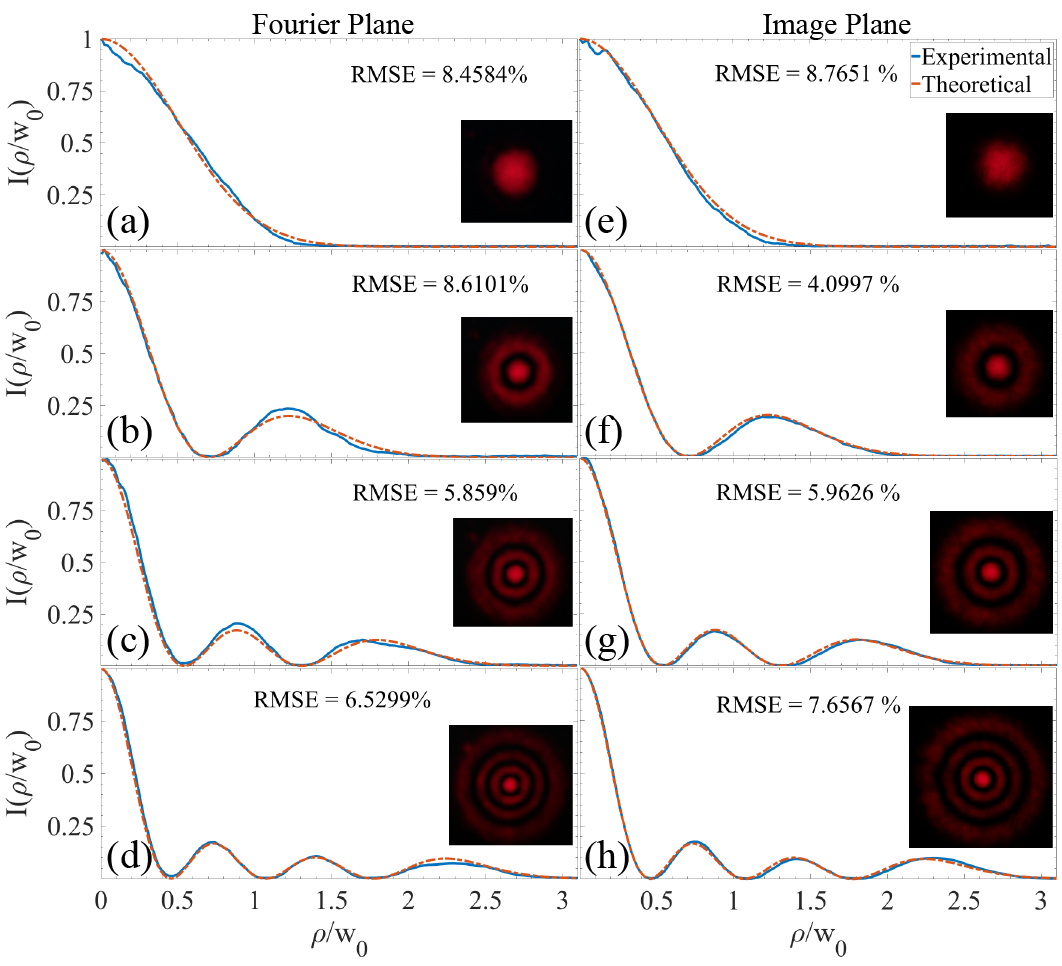}
	\caption{Azimuthally averaged intensity distributions of the generated radial modes at the (a-d) Fourier plane, and at the (e-h) image plane of SLM1 recorded in the experiment (blue), and from theory for radial modes $LG^0_0$, $LG^0_1$, $LG^0_2$ and $LG^0_3$ (from top to bottom). Insets show the field intensities recorded by a CCD camera. The goodness-of-fit is represented as the percent root-mean-square-error (RMSE) on the plots.}
	\label{fig:Fidelity}
\end{figure}

The Laguerre-Gaussian (LG) modes propagating along $z$ with waist size $w_0$ can be written in the position representation in cylindrical coordinates as \cite{Krenn}
\begin{align*}
LG^l_p (r, \phi, z) &= \sqrt{\frac{2p!}{\pi(p + |l|)!}} \frac{1}{w_z} \Bigg(\frac{\sqrt{2} r}{w_z} \Bigg)^{|l|} L^{|l|}_p \Bigg( \frac{2r^2}{w^2_z} \Bigg) \cdot \\
&  \cdot \exp \Bigg( - \frac{r^2}{w^2_z} + i\Big(l\phi + \frac{kr^2}{2R_z} - (2p + |l| + 1 )\phi_g\Big) \Bigg) .
\end{align*}
where $w_z = w_0 \sqrt{1 + \Big( \frac{z}{z_0}\Big)^2}$ is the waist size of the mode after propagating a distance $z$, $z_0 = \pi w^2_0/\lambda$ is the Rayleigh range, $R_z = z + z^2_0/z$ is the radius of curvature of the wavefront, and $\phi_g = \tan^{-1} {z/z_0}$ is the Guoy phase shift. The radial modes considered here have a zero azimuthal index, or $l = 0$.\\
\indent As a first step, the fidelity of the generated modes from SLM1 was verified. Without adding any correcting phases on the SLM1, the intensities of different $LG^0_p$ modes were captured by a CCD camera at the Fourier plane, as well as the image plane of SLM1. The mean intensity distributions were then obtained from the recorded intensities by averaging over the azimuthal coordinates. The average measured intensity distributions were then compared with the theoretical intensities obtained by taking the squared modulus of the field distribution of $LG^0_p(r, \phi, z=0)$.\\
\indent Figures~\ref{fig:Fidelity}(a-d) and (e-h) compare the averaged intensity distributions of $LG^0_p$ modes, obtained experimentally (blue) and theoretically (red), at the Fourier plane and at the image plane respectively. The intensity of the modes recorded on a CCD camera are shown in the insets. The root-mean-squared error between the theoretical cross-section and the cross-section obtained from the experiment is less than $10\%$ for both the image plane and the Fourier plane (as indicated in the inset), which confirms that the generated $LG^0_p$ modes are of high fidelity.
\subsection{Tables of coupling efficiencies}
Tables~\ref{table:calculated} and~\ref{table:experiment} show the calculated and the measured coupling efficiencies ($C_{p_{\rm{in}}, p_{\rm{screen}}}$) respectively, of the various $LG^0_{p_{\rm{in}}}$ modes for the different pairs of added phase-screens (denoted by the radial index $p_{\rm{screen}}$). The coupling efficiencies are plotted in matrix form as the two crosstalk matrices shown in figure~\ref{fig:crosstalk}(a) and ~\ref{fig:crosstalk}(b). The conditional probability ($Pc_{p_{\rm{in}}, p_{\rm{screen}}}$) corresponds to the probability of detecting the $LG^0_{p_{\rm{in}}}$ mode provided that the Hilbert space comprises $\{ LG^0_0, LG^0_1, LG^0_2, LG^0_3 \}$. The conditional probability for both simulations and the experiment is given by
\begin{equation}
Pc_{p_{\rm{in}}, p_{\rm{screen}}} = \frac{C_{p_{\rm{in}}, p_{\rm{screen}}}}{\sum_{p_{\rm{in}}}C_{p_{\rm{in}}, p_{\rm{screen}}}}
\end{equation}

\begin{table}[ht!]

		\begin{tabular}{c c c c}
			\hline \hline
			$p_{\rm{screen}}$ & $p_{\rm{in}}$ & Coupling Efficiency & Conditional Probability \\
			& & $C_{p_{\rm{in}}, p_{\rm{screen}}}$ & $Pc_{p_{\rm{in}}, p_{\rm{screen}}}$\\
			\hline 
			0 & 0 & 0.9947 & 0.994501 \\ 
			0 & 1 & 0.0001 & 0.00009998 \\ 
			0 & 2 & 0.0018 & 0.00179964 \\
			0 & 3 & 0.0036 & 0.00359928\\
			1 & 0 & 0.0000 & 0.0000  \\
			1 & 1 & 0.8310 & 0.996761\\
			1 & 2 & 0.0001 & 0.000119947\\
			1 & 3 & 0.0026 & 0.00311863\\
			2 & 0 & 0.0000 & 0.0000\\
			2 & 1 & 0.0025 & 0.0030499\\
			2 & 2 & 0.8170 & 0.996706\\
			2 & 3 & 0.0002 & 0.000243992\\
			3 & 0 & 0.0000 & 0.0000\\
			3 & 1 & 0.0173 & 0.0209468\\
			3 & 2 & 0.0004 & 0.00048432\\
			3 & 3 & 0.8082 & 0.978569\\
			\hline
		\end{tabular}
\caption{\label{table:calculated}%
	Table of calculated coupling efficiencies and conditional probabilities of the  $LG^0_{p_{\rm{in}}}$ modes for the pairs of phase-screens corresponding to radial index $p_{\rm{screen}}$.	}
\end{table}

\begin{table}[ht!]

		\begin{tabular}{c c c c}
			\hline \hline
			$p_{\rm{screen}}$ & $p_{\rm{in}}$ & Coupling Efficiency & Conditional Probability \\
			& & $C_{p_{\rm{in}}, p_{\rm{screen}}}$ & $Pc_{p_{\rm{in}}, p_{\rm{screen}}}$\\
			\hline
			0 & 0 & 0.30537 & 0.950736  \\
			0 & 1 & 0.00686248 & 0.0213656 \\
			0 & 2 & 0.00497067& 0.0154757\\
			0 & 3 & 0.00399011 & 0.0124228 \\
			1 & 0 & 0.00399687 & 0.0151793 \\
			1 & 1 & 0.245118& 0.93091\\
			1 & 2 & 0.00608885 & 0.0231242\\
			1 & 3 & 0.00810648 & 0.0307868\\
		    2 & 0 & 0.0111205 & 0.0538439 \\
		    2 & 1 & 0.0192632& 0.0932698\\
		    2 & 2 & 0.150514 & 0.728766\\
		    2 & 3 & 0.0256349 & 0.124121\\
			3 & 0 & 0.0153835 & 0.0718829\\
			3 & 1 & 0.00634927 & 0.0296683\\
			3 & 2 & 0.0480315 & 0.224437\\
			3 & 3 & 0.144244 & 0.674011\\
			\hline
		\end{tabular}
	\caption{\label{table:experiment}%
	Table of measured coupling efficiencies and conditional probabilities of the  $LG^0_{p_{\rm{in}}}$ modes for the pairs of phase-screens corresponding to radial index $p_{\rm{screen}}$.	}
\end{table}

\end{document}